\documentclass{emulateapj}

\newcommand{\HI}{H\,{\sc i}}
\newcommand{\HII}{H\,{\sc ii}}

\newcommand{\arcs}{\arcsec }

\slugcomment{Accepted to ApJL: 25 August 2008}

\shorttitle{Massive galaxy HIZOA J0836-43}
\shortauthors{Cluver et al.}

\begin{document}

\title{The hidden \HI-massive LIRG HIZOA J0836-43: Inside-out galaxy formation\altaffilmark{1}}

\author{M.E. Cluver\altaffilmark{2,5}, T.H. Jarrett\altaffilmark{3},  P.N. Appleton\altaffilmark{4}, R.C. Kraan-Korteweg\altaffilmark{2}, P.A. Woudt\altaffilmark{2}, B.S. Koribalski\altaffilmark{6}, J.L. Donley\altaffilmark{7}, K. Wakamatsu\altaffilmark{8} and T. Nagayama\altaffilmark{9}}

\altaffiltext{1}{This work is based on observations made with the {\it Spitzer} Space Telescope, which is operated by JPL/Caltech under a contract with NASA}
\altaffiltext{2}{Department of Astronomy, University of Cape Town, Rondebosch, 7700, South Africa}
\altaffiltext{3}{IPAC, California Institute of Technology, Pasadena, CA 91125}
\altaffiltext{4}{NASA Herschel S Center, California Institute of Technology, Pasadena, CA 91125}
\altaffiltext{5}{IPAC Visiting Graduate Fellow, California Institute of Technology, Pasadena, CA 91125}
\altaffiltext{6}{Australia Telescope National Facility, CSIRO, Epping, NSW 1710, Australia}
\altaffiltext{7}{Steward Observatory, University of Arizona, Tucson, AZ 85721}
\altaffiltext{8}{Faculty of Engineering, Gifu University, Gifu 501-1193, Japan}
\altaffiltext{9}{Department of Astrophysics, Nagoya University, Chikusa-ku, Nagoya 464-8602, Japan}

\begin{abstract}

HIZOA J0836-43 is an extreme
gas-rich ($M_{\rm{HI}}$=7.5$\times10^{10}\, M_{\sun}$) disk galaxy which lies hidden behind the strongly obscuring Vela region of the Milky Way. Utilizing observations from the {\it Spitzer Space Telescope}, we have found it to be a luminous infrared starburst galaxy with a star formation rate of $\sim 21\, M_{\sun}\, \rm{yr^{-1}}$, arising from 
exceptionally strong molecular PAH emission ($L_{7.7\micron} = 1.50 \times 10^{9} L_{\odot}$) 
and far-infrared emission from cold dust. The galaxy exhibits a weak mid-infrared continuum compared to other starforming galaxies and U/LIRGs. This relative lack of emission from small grains suggests atypical interstellar medium conditions compared to other starbursts. We do not detect significant $[$Ne\,{\sc v}$]$ or $[$O\,{\sc iv}$]$, which implies an absent or very weak AGN.
The galaxy possesses a prominent bulge of evolved stars and a stellar mass of 4.4($\pm$1.4)$\times10^{10}\, M_{\sun}$. With its plentiful gas supply and current star formation rate, a doubling of stellar mass would occur on a timescale of $\sim$2 Gyr. Compared to local galaxies, HIZOA J0836-43 appears to be a ``scaled-up" spiral undergoing 
inside-out formation,
possibly resembling stellar disk building processes at intermediate redshifts.

\end{abstract}


\section{Introduction}

Understanding the fundamental origin and formation of galaxies requires the synthesis of multi-wavelength observations and cosmological-based numerical simulations. Advances in both areas have created a compelling view of a cold dark matter dominated universe where structure forms hierarchically \citep{Wh78}. Of fundamental importance to this formalism is insight into the formation and evolution of galaxy disks \citep{Str07}.

Massive, gas-rich disk galaxies ($M_{\rm{HI}} > 10^{10} M_{\sun}$) are considered indicative of relatively unadvanced star building
systems, making them ideal laboratories for testing this formalism.
However, such systems are rare in the local universe and nearly all massive \HI-rich disk galaxies are inactive or only passively forming stars \citep{Spray95}, thus providing 
few clues as to their formation and evolution. For example, Malin 1, an extreme case of a giant low surface brightness galaxy \citep{Imp89}, has a 
dormant
star formation rate of $\sim 0.38\, M_{\sun}\, \rm{yr^{-1}}$ \citep{Rah07}.
In this letter, however,
we present evidence of an \HI-massive disk galaxy, HIZOA J0836-43, undergoing a vigorous starburst.
 
This galaxy, discovered as part of a blind \HI\ survey of the southern Zone of Avoidance,
contains 7.5$\times10^{10}\, M_{\sun}$ of \HI\ gas, has a total dynamical mass of 1.4$\times10^{12}\, M_{\odot}$ and a 20-cm derived star formation rate of $\sim$35\,$M_{\sun}\, \rm{yr^{-1}}$ \citep{Don06}. It also appears to have a prominent bulge in the near-infrared -- bulge-to-disk ratio of $\sim 0.80$ in the $K_{s}$ band -- central to an enormous, rapidly rotating \HI\ disk. 

The galaxy is located at $l$=262.48$\degr$, $b$=-1.64$\degr$, lying behind the Vela Supernova Remnant 
of the Milky Way. At optical wavelengths it is largely hidden by foreground gas and 
dust. We have therefore conducted a detailed infrared study of HIZOA J0836-43, using imaging and spectroscopy from the {\it Spitzer Space Telescope}, to fully reveal its morphology and past and present evolutionary state. Here we highlight key results from this study which provide new insights for understanding massive galaxy formation and evolution.
The adopted distance of HIZOA J0836-43, $D_{L}$=148 Mpc, is from \citet{Don06}, as derived from its recessional velocity, $v_{hel}$=10689 km\,$\rm{s^{-1}}$.

\section{Observations and Data Reduction}

\begin{figure*}[!t]
\begin{center}
\includegraphics[width=7cm]{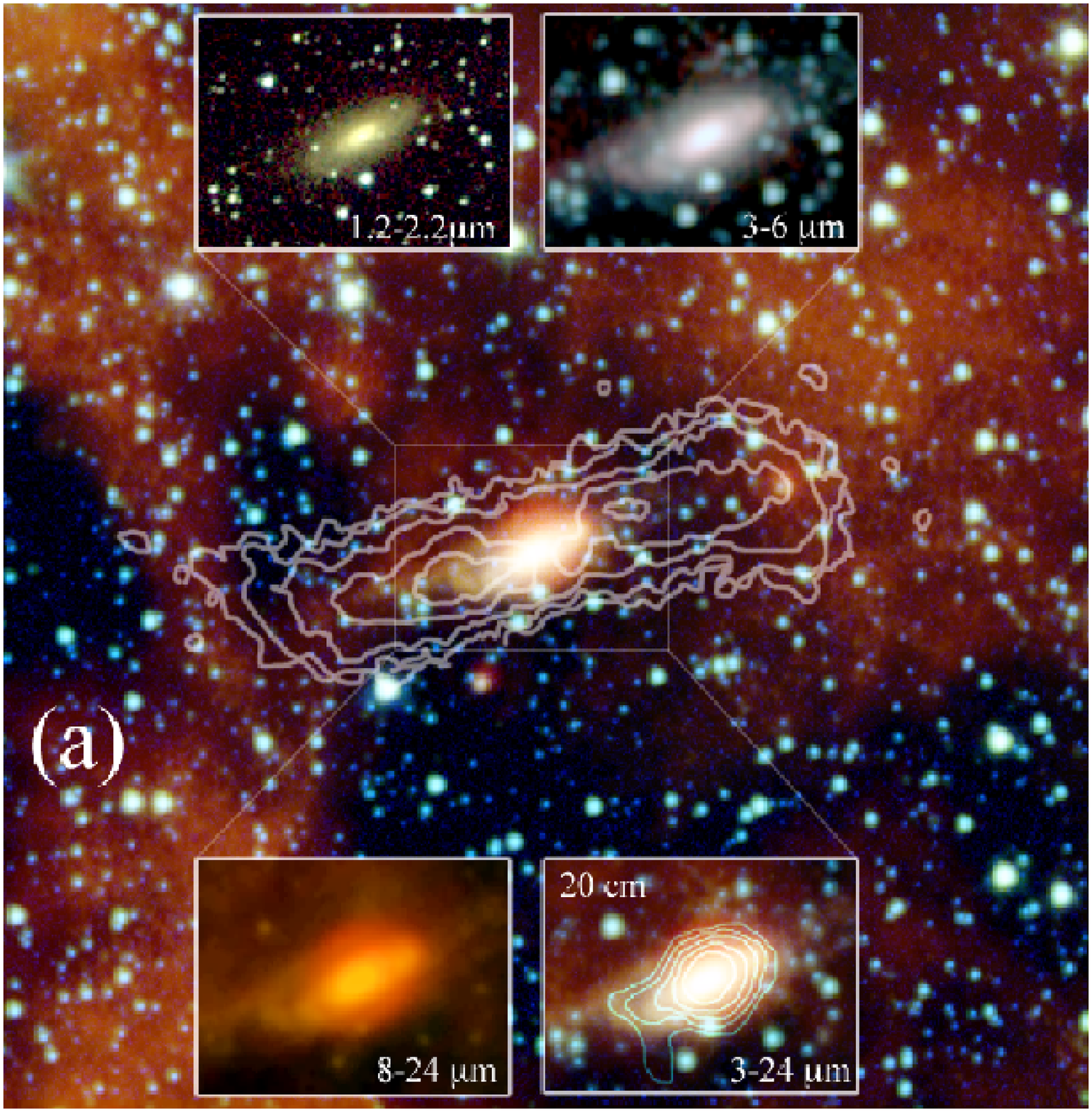}
\includegraphics[width=7cm]{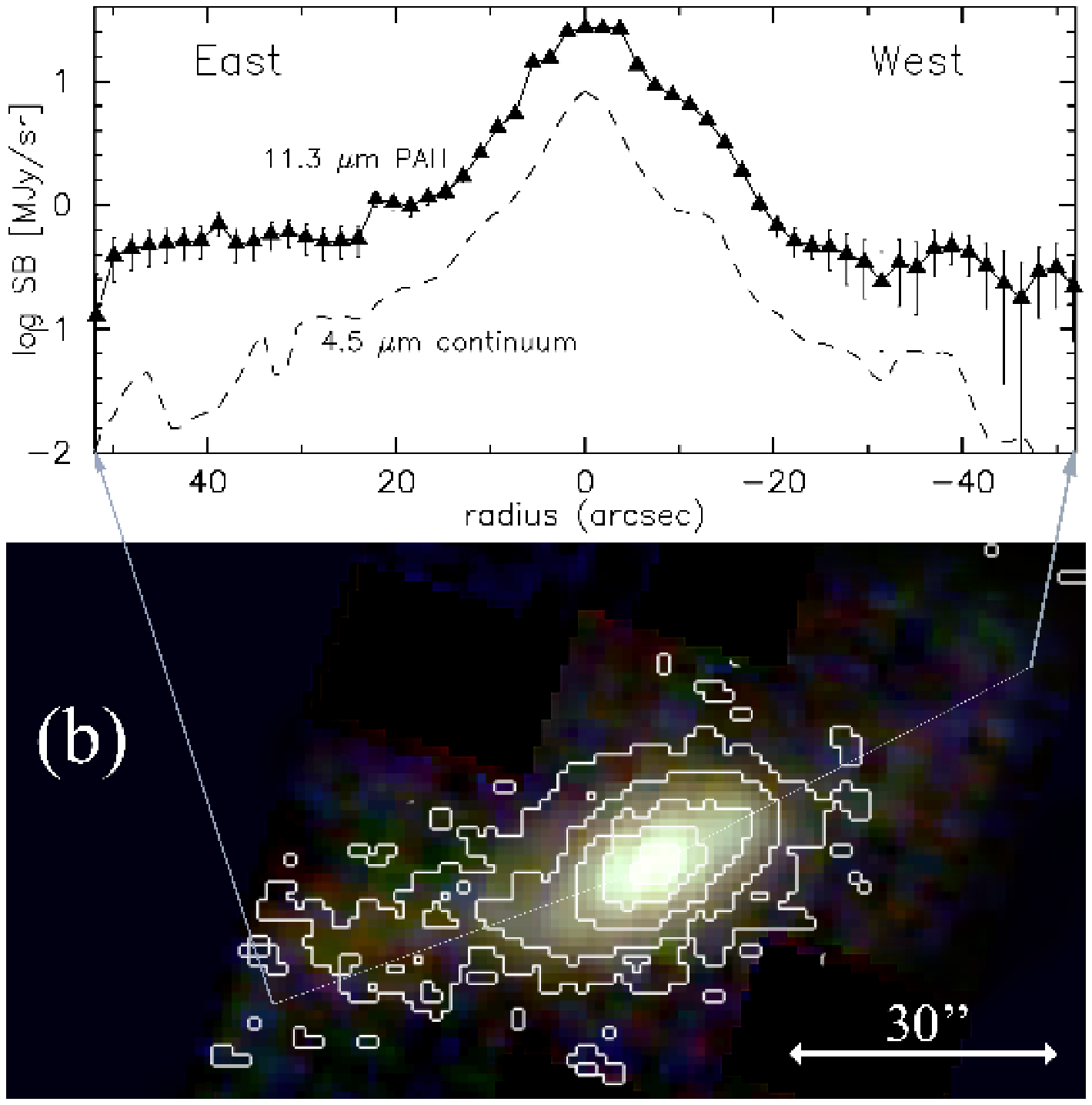}
\caption[Composite images of HIZOAJ0836-43]
{\small{
Infrared view of HIZOA J0836-43 
through the 
Vela region of the 
Milky Way.
(a) Composite 1 - 24$\mu$m image of HIZOA J0836-43
( $\sim$4\arcmin\ FOV $= \sim$170 kpc)
with \HI\ contours overlaid; insets show zoomed versions
of the inner 1\arcmin\ region; 
20-cm radio contours overlay
the 3 - 24$\mu$m inset; 
(b) Composite of spectral maps for the 6.6$\mu$m (blue), 7.7$\mu$m (green) and 11.3$\mu$m (red) PAH emission bands, overlaid with $[$Ne\,{\sc ii}$]$\,12.8$\mu$m contours; the plot shows the 11.3$\mu$m PAH and IRAC 4.5$\mu$m surface brightness across the East-West major axis,  demonstrating the East warp and
disk extended PAH emission relative to the bulge-dominated stellar light.
}}
\label{fig:1}
\end{center}
\end{figure*}

Near-infrared (NIR) and mid-infrared (MIR) imaging and spectroscopy served as primary data sets to study HIZOA J0836-43.  From the ground, 
simultaneous $J H K_{s}$ images were obtained in 2006 April using the 1.4m InfraRed Survey Facility (IRSF) and the 0.45\arcs\ pixel scale SIRIUS camera \citep{Nag03}, 
achieving an angular resolution of $\sim$1\arcs.
The {\it Spitzer Space Telescope} was used to obtain imaging \citep{Faz04,Riek04} and spectroscopy \citep{Hou04} in 2007 April and May. 
IRAC (3.6 - 8\micron) achieves a spatial resolution of $\sim$2\arcs\ for all bands, and 
MIPS (24, 70, 160\micron) $\sim$6\arcs, 18\arcs\ and 40\arcs\ for the 24, 70, 160 $\mu$m, respectively.
Primary data reductions were done by the {\it Spitzer} Science Center (SSC) science pipeline (version S16.1.0) and using the SSC-developed MOPEX tool to produce final science-grade images. Galaxy photometry was performed using a matched elliptical aperture. The aperture was determined using the IRAC-3.6\micron\ image, the optimal 
window to determine the shape and size of the galaxy after factoring in sensitivity, angular resolution and foreground extinction. A symmetric isophotal fit of the light distribution down to the 1$\sigma$ sky level of this band was performed; the resulting aperture has a semi-major axis radius of 39.4\arcs, axis ratio of 0.42 and a position angle of $\sim$110$\degr$. Foreground contaminating stars were masked from all images and replaced by the corresponding isophotal value of the source. The local background was determined from the median pixel value distribution within a surrounding annulus. 
The 
formal photometric uncertainties are $\sim$5\%, and $\sim$10 -- 20\% for the NIR/IRAC and MIPS calibration error, respectively.

IRS spectroscopy was obtained using the Short-Low (SL; $5-14\micron$), Short-High (SH; $10-20\micron$) and Long-High (LH; $19-38\micron$) modules. Integration times of 3$\times$60s were used for SL mapping (R$\sim64 - 128$), and 4$\times$30s and 4$\times$14s for SH and LH (R$\sim$ 600) in staring mode.
The data were first processed through the SSC S16.1.0 
pipeline. The SL observations consisted of three separate mappings: Center, East and West of the galaxy nucleus, each covering 0.4\arcmin$\times$0.7\arcmin\ with $<$10\% overlap between adjacent maps. SH/LH observations were centered on the nucleus, and for background subtraction, a region $\sim$1\arcmin\ south of the galaxy, not confused by foreground Galactic emission, was used. Spectral cubes and corresponding spectra were extracted using CUBISM \citep{Smi07}. A 9.25\arcs\ aperture was used for the nucleus (SL, SH, LH) and East and West disks (SL), and a 37\arcs\ aperture for the entire galaxy (SL).

\section{Results}

\subsection{Imaging and Photometry}

Fig. 1a shows a composite $1-24\micron$ image of the galaxy and its local environment ($\sim$4$\arcmin$). The \HI\ observations \citep{Don06}, also shown, demonstrate the enormous diameter ($\sim$3$\arcmin = \sim$130 kpc) and \HI\ mass ($M_{\rm{HI}}$=7.5$\times10^{10}\, M_{\sun}$).  The $1-5\micron$ window traces the evolved stellar population, while the MIR is sensitive to the interstellar medium:  thermal dust continuum and emission from PAH (polycyclic aromatic hydrocarbon) molecules.
PAHs produce broad emission bands in the MIR and are linked to ongoing or recent star formation \citep{All85}. 
The emission likely arises from photodissociation regions (PDRs) which form adjacent to \HII\ regions produced by star formation \citep{Holl97}. The 20-cm radio continuum (Fig. 1a) is closely correlated with the $8-24\micron$ emission, indicating a common star-formation origin. 
The infrared emission is clearly extended and exhibits asymmetry, or warp, along the eastern side in the PAH spectral map and surface brightness distribution along the major axis (Fig. 1b).

\begin{table}[!t]
{\scriptsize

\caption{Infrared Photometry of HIZOA J0836-43 \label{tab:res1}}
\begin{tabular}{p{1.25cm} p{0.8cm} p{0.5cm} p{0.8cm} p{0.8cm} p{0.8cm} c}
\tableline
\tableline
\\[0.25pt]
  Band      &  ${\lambda}$  &  $A_{\lambda}$   &  $F_{\nu}$  &   ${\nu} L_{\nu}\,\,^a$  &  $\rm{r_{eff}}$  &  \rm{SB}(${\rm{r_{eff}}}$)\\
           &  $({\mu}$m)    &   (mag)   &    (mJy)               &  ($10^{9} L_{\odot}$)   &  (arcsec)  & $^{b}$   \\

\tableline
\\[0.25pt]

$I\,\,^a$      &      0.82   &  4.2   & 28.12   &  69.2  &  ---  &  ---
\\
$J$         &   1.25  &  2.0   & 53.18  &  85.6  &  8.49  &  16.75
\\
$H$        &    1.63  &  1.3  & 60.53   &   73.1  &  7.88  &  16.03
\\
$K_{s}$    &    2.14  & 0.8   &  52.17  &  49.0   &  8.00  &  15.74
\\
IRAC-1    &     3.53  &  0.4  & 31.40 &    17.6   &  8.94   &  15.56
\\
IRAC-2     &    4.46  &   0.3    & 21.72  &   9.74  &  8.14  &  15.33
\\
IRAC-3     &    5.67  &  0.3  & 45.20  &    15.7  &  7.34   &  13.70
\\
IRAC-4     &   7.70   &  0.3   & 145.31   &     36.7 &   7.34  &  11.74
\\
MIPS-1    &   23.7  & ---  & 126.52  &     10.6  &   ---   &  ---
\\
MIPS-2    &   71   & --- &  2291.3   &    65.9   &  ---   &  ---
\\
MIPS-3   & 156   & --- & 3082.1   &   38.7   &  ---   &  ---
\\
Total IR$\,^c$   & 3-1100 &    &  &  119.5  
\\
\tableline
\end{tabular}
\tablecomments{Aperture parameters: $a=39.35\arcs, b/a=0.42, \phi=-70\degr$.
All measurements have been corrected for foreground dust and internal extinction; $A_{V}$ = 7.3. IRAC and MIPS measurements have been aperture corrected as follows:
0.940, 0.974, 0.871, 0.814, 1.107, 1.240, 1.705, for 3.6, 4.5, 5.8, 8.0, 24, 70, 160$\mu$m respectively.
$^a$\,\,$D_{L}$=148 Mpc from Donley et al. (2006);
$^b$ (mag arcsec$^{-2}$);
$^c$ Using the relation of Dale and Helou (2002).
\\
}

}
\end{table}

This extended morphology, also evident in the radio continuum, is reminiscent of a tidal tail, possibly due to a minor disturbance.
At the Galactic location of HIZOA J0836-43 ($l$=262.48$\degr$, $b$=-1.64$\degr$) there is severe foreground dust obscuration. In order to estimate the Galactic extinction, we exploit the morphology-independent NIR colours of galaxies \citep{Jar00}, combined with the NIR sensitivity to the relatively well-modeled 
stellar population, by comparing the galaxy Spectral Energy Distribution (SED) to dust-reddened templates for Population II-dominated galaxies \citep[GRASIL code;][]{Sil98}.  
The best-fit SED corresponds to a well-constrained extinction of $A_{V}$=7.3$\pm$0.2 mag using the \citet{Car89} extinction law convolved with the IRSF and IRAC bandpass filters.
Table 1 presents the extinction-corrected global photometry and central surface brightness of the galaxy. The resulting SED is presented in Fig. 2a. We include GRASIL templates for E, S0 and Sc types for comparison, as well as the spectrum of M82 \citep{Stu00}, the prototypical local starburst galaxy.  
The SED shows that in the MIR the galaxy resembles a Sc-type galaxy with strong emission from a dust continuum and PAH molecules.
We see strong FIR emission longwards of 60\micron\ (see Table 1), indicating a significant cold dust component that dominates the total IR luminosity, $L_{\rm{TIR}}$=$1.2\times{10}^{11}L_{\odot}$, giving rise to a luminous infrared Galaxy (LIRG).
Fitting a two-component modified blackbody curve ($\beta$=1.5) to this FIR continuum, we estimate a cold dust temperature of $\sim$30K. 
Assuming negligible contribution from AGN, we estimate a total SFR of $\sim$21\,$M_{\sun}\, \rm{yr^{-1}}$ based on our global MIR measurements and the relation of \citet{Ken98}. This value is consistent with that of local LIRGS \citep{Wan06}, but much higher than that of typical disk spirals \citep{Kew02}.

\subsection{MIR Spectroscopy}

\begin{figure}[!t]
\includegraphics[width=8.5cm]{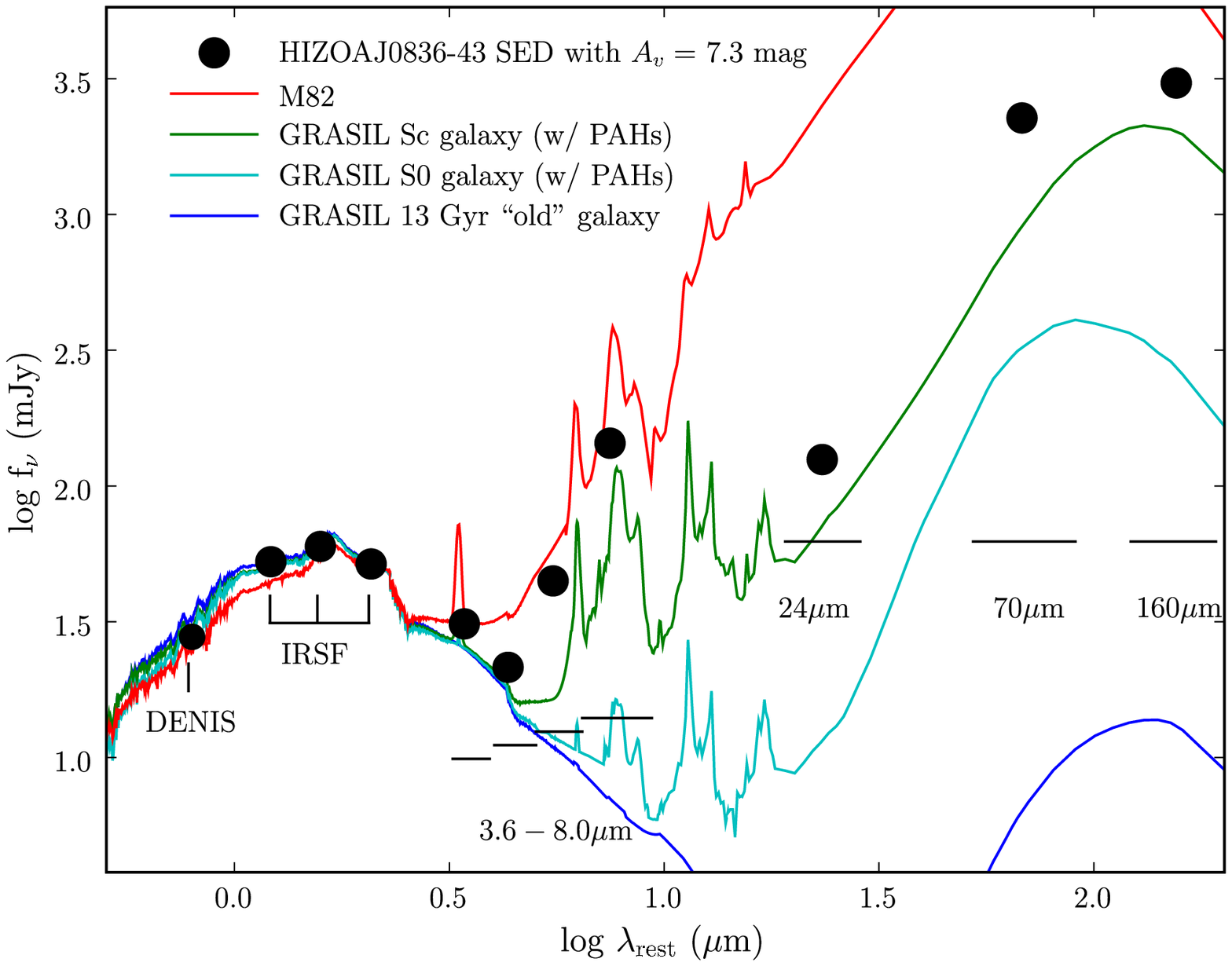}
\\
\\
\\
\includegraphics[width=8.5cm]{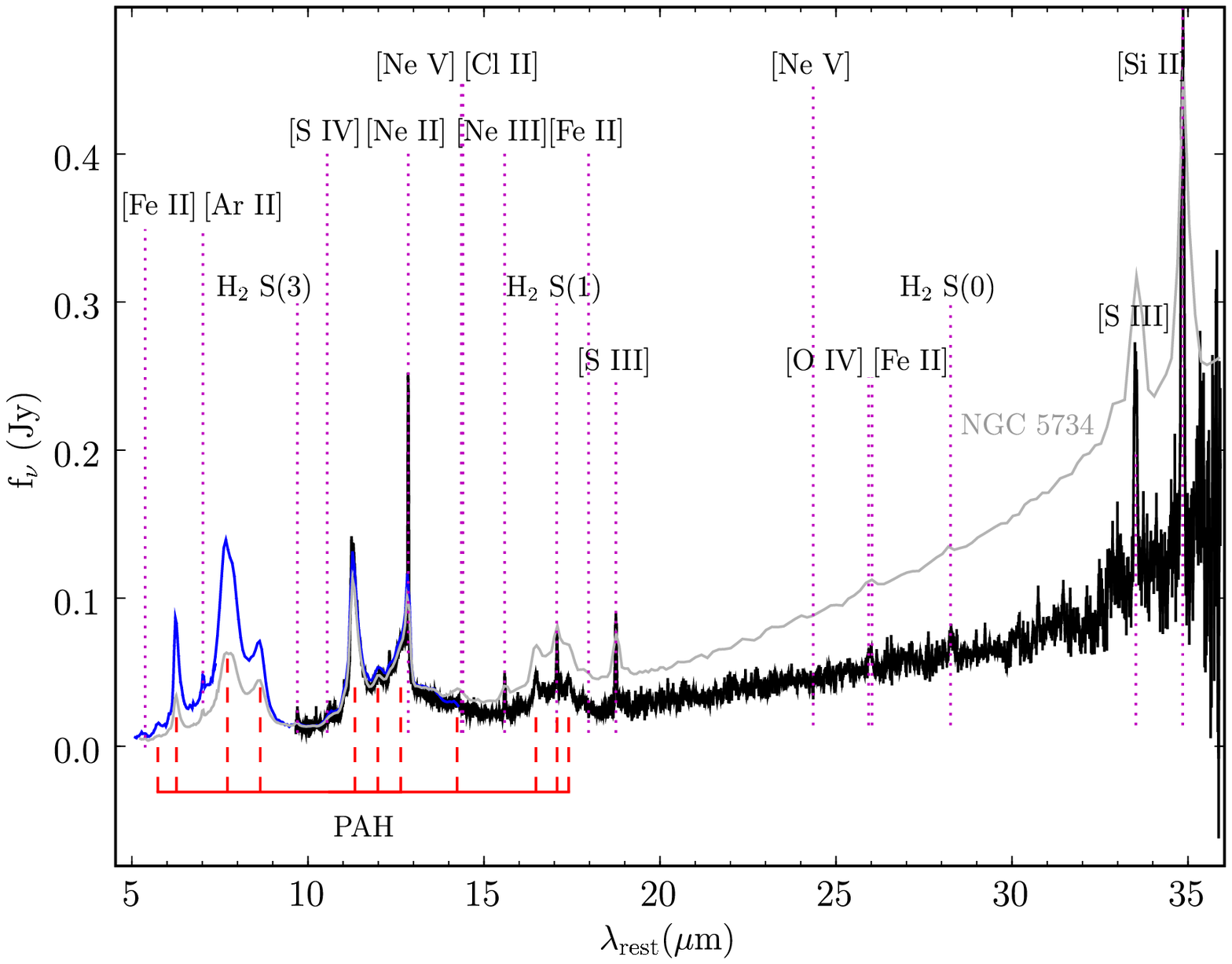}
\caption[Massive galaxy SED and Spectrum]
{\small{{a) Global SED with model templates and M82 spectrum for comparison. The templates are normalised to the galaxy $K_{s}$ band value.  b) Combined low resolution (blue) and high resolution (black) spectrum for the massive galaxy nuclear region. 
For comparison, we show the SL/LL spectrum of the similarly cold GOALS galaxy NGC 5734 (grey).}
}}
\label{fig:2}
\end{figure}

The MIR spectroscopy of HIZOA J0836-43 is dominated by strong PAH emission that peaks in the nuclear region and appears extended compared to the distribution of stellar light, as traced by the 4.5\micron\ emission (Fig 1b).
The nuclear light is shown in Fig. 2b, which combines the SL, SH and LH spectra, $5-36\micron$. The 6.2 and 7.7\micron\ PAH features have equivalent widths of $0.72 \mu \rm{m}$ and $0.82 \mu \rm{m}$ respectively. This is $\sim$20\% and 50\% respectively, larger than the average value (0.53\micron) found by \citet{Bra06} for starburst galaxies. We attribute this to a combination of power in the PAH emission lines and the absence of a sharply rising dust continuum typical of starburst systems 
(e.g., M82 in Fig. 2a). The galaxy lies on the relation between $L(6.2\micron)$ and $\nu L_{\nu}(5.5\micron)$ found for starburst galaxies \citep[see Fig. 8 of][]{Des07}, but at the high-luminosity end. 
However, the flux density continuum ratio $f_{\nu}(6\micron)/f_{\nu}(15\micron)$ of $\sim$1 is indicative of a quiescent disk galaxy \citep{Dal00}.  Similarly, the rotational lines of $\rm{H_{2}}$, a direct tracer of the warm molecular component, are relatively weak in the galaxy.
Using the relative excitations of the {\it S}(1), {\it S}(2) and {\it S}(3) lines, we estimate a warm molecular hydrogen temperature of T$\sim$330K and a mass of only $\sim$1.3$\times10^{7}\, M_{\sun}$. 
Yet, the fine-structure lines of Ne, S, Si and Ar are all relatively strong; e.g., 
the $[$Ne\,{\sc iii}$]\, 15.56 $/$[$Ne\,{\sc ii}$]\,12.81$ ratio, which provides an indication of the hardness of the radiation field, 
is $\sim$0.11,  comparable to typical values for starbursting systems \citep{Bra06}. We do not detect significant $[$Ne\,{\sc v}$]$ or $[$O\,{\sc iv}$]$ which implies an absent or very weak AGN. Spectroscopy results are discussed in more detail in \citet{Cluver08}.

\section{Discussion}

For a galaxy in the local universe HIZOA J0836-43 possesses a number of unusual infrared properties; considered in combination with the massive reservoir of gas that feeds it, this could be a rare instance of a local galaxy undergoing inside-out evolution, possibly resembling galaxy formation at earlier epochs.  Here we compare properties of HIZOA J0836-43 with those of local and intermediate redshift samples.

The paucity of warm dust is evident from the weakly rising continuum seen in its spectrum (Fig. 2b), consistent with its low $L_{24\micron}/L_{70\micron}$ colour compared to normal and star-forming systems, e.g., as compared to both the SINGS ({\it Spitzer} Infrared Nearby Galaxy Survey) sample, and the relatively nearby Great Observatories All-sky LIRG Survey (GOALS). 
In contrast, the strength of the PAH emission in HIZOA J0836-43, both in luminosity and relative strength of the bands compared to the continuum, is amongst the largest observed in any star-forming galaxy \citep{Peet04}, implying unusually strong PDR emission. 
These unusual MIR properties may arise from (1) a paucity of very small grains (VSGs) and/or (2) a soft UV radiation field.  Powered by massive star formation, transiently heated VSGs are thought to be the source of MIR radiation.  A weaker radiation field would give rise to cooler VSGs \citep{Dra07}, which would re-radiate at longer FIR wavelengths along with the large grain, cool (T $\sim$20 K) component. Both mechanisms would be consistent with an evolutionary phase in which the activity is confined to heavily gas/dust-obscured star forming regions, thus shielding the hard UV radiation from the rest of the disk, which in turn would yet to have experienced significant grain processing. 

\begin{figure}[!t]
\begin{center}
\includegraphics[width=8.5cm]{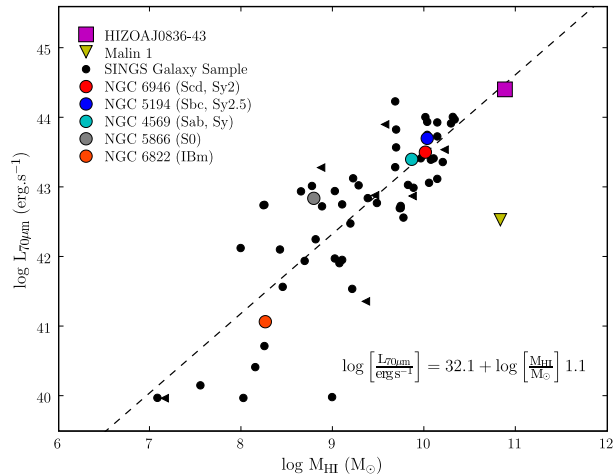}
\caption[Comparison with SINGS and Malin 1]
{\small{Infrared luminosity vs. \HI\ mass, comparing HIZOA J0836-43, Malin 1 and the SINGS galaxy sample.
For Malin 1 (yellow triangle) a 70\micron\ upper limit \citep{Rah07} is shown;  
upper limits for \HI\ mass 
are shown as black triangles.}
}
\label{fig:3}
\end{center}
\end{figure}

The growth progress of HIZOA J0836-43 is deduced from the stellar bulge population and current star formation rate. 
We estimate a stellar mass of 4.4($\pm$1.4)$\times10^{10}\, M_{\sun}$ for the galaxy \citep[from the relation of][]{Bell03}
and hence a specific star formation rate, SFR per stellar mass, of $\sim$0.5 $\rm{Gyr}^{-1}$. 
Compared to local LIRGS \citep[see Fig. 5 of][]{Wan06}, this implies active stellar building 
as facilitated by its plentiful supply of gas, with a doubling of stellar mass in $\sim$2 Gyr.
The specific star formation rate of the galaxy in combination with its stellar mass appears typical for star-forming galaxies at $z\sim 0.7$, where gas fractions of disks were likely higher compared to local galaxies \citep{Bell05, Per05}.
Fig. 3 shows that the MIR luminosity is strongly correlated with \HI\ content; 
even with its extreme \HI\ mass, HIZOA J0836-43 appears consistent with being a ``scaled-up" disk galaxy, unlike Malin 1 which is explicitly quiescent by comparison. This suggests relatively ``normal'' evolution in HIZOA J0836-43, despite lying at the extreme high end (i.e., early evolutionary stage) of the relation.

HIZOA J0836-43 is a gas-rich spiral galaxy exhibiting a warp in its disk (Fig. 1a), likely the result of a disturbance in its recent past
($<<$ 1 Gyr). 
Such an event could cause the observed starburst as gas from the extended \HI\ disk flows into the central region of the galaxy. Hence the starburst is powered by gas consumption, as opposed to a major merger event. Gas-rich galaxies, like HIZOA J0836-43, were likely more common in the distant universe and there is evidence that gas consumption and not merger interactions were driving stellar mass growth in the distant universe \citep{Dad08}. Recent work has suggested that many LIRGs seen at intermediate redshifts ($z\sim0.8$) achieve heightened star formation as a result of the high gas fractions of their disks and were less dependent on major interactions, compared to local LIRGs, to induce starburst activity \citep{Mel08, Mar06}.

Observational evidence suggests that disk galaxies evolve along the stellar mass-radius relation and have built stellar mass intensely since $z=1$, on average by means of inside-out growth \citep{Bar05,Tru06}. 
Comparing the extended regions of active star formation with the more centrally concentrated
stellar bulge distribution (e.g., Fig 1b radial profile) suggests that HIZOA J0836-43 is undergoing vigorous disk building,
an instance of inside-out growth.
This combined with its PDR-dominated emission manifested as strong PAH emission coupled with a weak MIR continuum, makes it enigmatic in the local universe. Observing a galaxy at such a key point in its evolution could have far reaching implications for theories of galaxy formation and evolution.

\acknowledgements

We thank D. Dale and SINGS, J. Howell and GOALS for data access.
We are grateful to S. Carey, G. Helou, S. Lord, J. Mazzarella and B. Madore for insightful discussions.
Support for this work was provided by NASA through an award issued by JPL/Caltech. MC, RKK and PAW thank the NRF for financial support. MC thanks IPAC/Caltech for financial support through a Visiting Graduate Fellowship.

\end{document}